\documentclass[a4paper,11pt]{amsart}
\usepackage{multicol,ifthen}
\usepackage{amssymb,stmaryrd}
\usepackage{harvard}
\usepackage{graphicx}

\newcommand{\beq}{\begin{equation}}
\newcommand{\eeq}{\end{equation}}
\newcommand{\beqarr}{\begin{eqnarray}}
\newcommand{\eeqarr}{\end{eqnarray}}
\newcommand{\beqa}{\begin{eqnarray*}}
\newcommand{\eeqa}{\end{eqnarray*}}
\begin{document}

\title{Einstein universes  stabilized }
\author[E. Scholz]{Erhard Scholz$\,^1$}
\footnotetext[1]{\, scholz@math.uni-wuppertal.de\\
\hspace*{7mm}University Wuppertal,  Interdisciplinary Center for Science and 
Technology Studies\\
\hspace*{6.5mm} and Department C: Mathematics and Natural Sciences } 

\date{September 18, 2007 }

\begin{abstract}
The hypothesis that gravitational self-binding energy may be the  source for the vacuum energy term of cosmology is studied in a Newtonian Ansatz.  For spherical spaces the attractive force of gravitation and the negative pressure  of the vacuum energy term  form a  self stabilizing system under very reasonable restrictions for the parameters, among them a characteristic coefficient $\beta$ of self energy. In the Weyl geometric approach to cosmological redshift, Einstein-Weyl universes  with observational restrictions of the curvature parameters are dynamically stable, if $\beta$ is about $40 \%$ smaller than in the exact Newton Ansatz or if the space geometry is  elliptical.
\end{abstract}

\maketitle 

\section{Introduction}
The physical nature of the cosmological vacuum is a terrain wide open  for questions. They have become increasingly pressing, since  observational data are indicating an important role for the vacuum contribution to the total balance of the Einstein equation in realistic cosmological models. 

Cosmic vacuum is characterized by a thermodynamical neutrality in the following sense:  The expansion work of a small vacuum volume during  space expansion has to compensate the increase of the energy content while enlarging  the  volume. The characteristic relation 
$ p = - \rho_{vac} $ (setting $c=1$)
between vacuum pressure $p$ and the vacuum energy density $ \rho_{vac}$ follows from this property. The energy momentum tensor of the vacuum $ T^{vac}$ (denoting  tensors of the type  $T=(T_{ij})$ by their coordinate free expression $T$)   therefore satisfies the relation
$ T^{vac} = \rho_{vac} \, g \, $
  with $g=(g_{ij})$ the Lorentz metric of spacetime; in coordinate expressions
\[T^{vac}_{ij} = \rho_{vac} \, g_{ij} \; .  \]
$T^{vac} $  has the {\em form} of an energy momentum tensor derived from a cosmological constant term in the Lagrange action,  if $\rho_{vac}$ is supposed constant. 

Recently H.-J. Fahr e.a. have proposed  considering gravitational self-binding energy of cosmic matter distributions,  and its density $\rho_{grav}$,  as a possible source of  vacuum energy  \cite{Fahr/Overduin,Fahr:BindingEnergy}
\beq \label{vacuum energy}  \rho_{vac}  = \rho_{grav}  \; .\eeq
In this case  vacuum energy density can {\em  no longer be considered constant}, like in its characterization by a classical cosmological constant ($\Lambda $-) term. It rather  will be dependent on the matter-energy content of the universe. A similar ideas has been proposed earlier  by \cite{Fischer:1993}.

This is a very welcome modification of the received approach.
Fahr and Heyl give a completely convincing argument, why this should be so:\\
``A constant vacuum energy doing an action on space by accelerating its expansion, without itself being acted upon, does not seem to be a concept conciliant with basic physical principles'' \cite[2]{Fahr:BindingEnergy}. \\
Present cosmological models often assume a cosmological constant approach and are subject to the verdict of this simple basic  criticism.

It is an open problem how to characterize gravitational self-binding energy in cosmological models. Fahr and Heyl use an approach with a Poisson equation for the cosmic potential with respect to radial coordinates (ibid., equ. (16)). They consider an exchange between vacuum energy and mass energy in both directions,  mass creation from vacuum energy in the one direction and vacuum energy induced from self energy of gravitating masses in the other. From this they derive a new cosmological model which they call the {\em economic universe}. 

As matter density on the cosmological level is extremely small, it is not per se nonsensical to investigate Newtonian approximations for  potential and self-binding energy of a homogeneous mass distribution. Although  the potential of a homogenous mass distribution in euclidean space is infinite, it  may be finite in closed  spaces like the 3-sphere $S^3$ or spherical spaces finitely covered by it. Geometrically most interesting is the positively curved space of  non-euclidean geometry, arising from $S^3$ by antipodal identification. In classical geometrical language  it is the  {\em elliptical space} $\mathcal{E}^3$ or,  in more recent terminology, the {\em round projective 3-space} with the metric inherited from the classically metricized 3-sphere.  It thus seems worthwhile to study Robertson-Walker cosmologies with closed spacelike fibres and their dynamics under the assumption of a variable vacuum  term given by the  gravitational self-binding energy in Newtonian approximation.  

In the sequel we study the densities of gravitational self-binding energy of homogeneous mass distributions for the two most simple spherical spaces, $S^3$ and $\mathcal{E}^3$ (section 2). It will be shown that it differs only by a typical factor $\beta$ in the expression
\[ \rho_{vac} = \beta \, G_N \rho_{tot}^2 f^2 \; , \]
where $G_N$ denotes the Newton constant and $f$ the scaling function of the sperical space.
  Interestingly the total energy density $\rho_{tot}$  and vacuum energy show different scaling behaviour 
\[ \rho_{tot} \sim   f^{-3} \; ,\quad \quad  \rho_{vac} \sim  f^{-4} \; . \]
 Because of the different fall off of the attractive term $\rho_{tot}$   and the expansive term  $p=- \rho_{vac}$   solutions close to the static Einstein universe are dynamically stable. The simplified Rauychaudhury equation has  a Ljapunov stable neighbourhood of the Einstein universe, while outside certain bounds an unlimited expansion occurs (section 3). 

Of course, non-expanding cosmological models may acquire physical meaning only if cosmological redshift has a field theoretic origin rather than ``space expansion'', which is just another view of a time dependent spatial metric. Weyl geometry allows an intriguing geometric characterization of such an assumption. Moreover, Weyl geometric versions of Einstein universes have good empirical properties. Their parameter determined by supernovae data  corresponds to a value $\beta  \approx 2$ for the characteristic coefficient of the gravitational energy, mentioned above. This   value is consistent with a Lyapunov stable regime of the $\mathcal{E} ^3$, while it leads to instability in an exact Newtonian Ansatz for the gravitational binding energy in the  sphere $S^3$ itself. Empirical tests of the $\mathcal{E}^3$ hypothesis, or for other spherical spaces, can be designed by studying symmetry constellations of quasars close to the redshift of the `cosmic equator' in the Einstein-Weyl universe (section 4).  

We draw the conclusion that already a Newton approximated gravitational
self-binding approach to the vacuum term leads to a vindication of  slightly generalized  Einstein universes as  dynamically stable and empirically interesting models  (section 5).

%\newpage
\section{Gravitional self-binding energy in spherical spaces}
For a heuristic motivation let us consider discrete masses $m_i$ of the same amount $m$ distributed in euclidean space on the nodes of a  bounded, cubically symmetric lattice with edge $a = a_0$ ($i \leq N$). Every two mass elements attract each other by Newton's law. Increasing  the lattice parameter   to $a = a_0 + da$ requires a finite work $dE$. The (absolute value) of the self-binding energy is given by expansion to infinite lattice edge lengths $a \rightarrow \infty$, summarily written:
\[  E(a _0) = \int_{a_0}^{\infty} dE  \; \]
This allows to determine self-binding energy of the lattice.  For an (unbounded) lattice extending over the whole space it is infinite. 

This is different in a spherical space. In the  sequel we first  consider the 3-sphere itself and calculate the self-binding energy of  homogeneously distributed discrete masses of averaged gross mass  density  $\rho$  (respectively, for $c=1$, energy density) in a continuity approximation according to Newton's law. Actually we assume large ``empty'' bits of space  between discretely distributed masses, carrying the corresponding gravitational field. Like in nuclear physics, where binding energy is emitted to the (electromagnetic) interaction field, the   self-binding energy set free by the gravitationally interacting masses is assumed to be distributed homogeneously in the vacuous space parts between the masses. Its density will be denoted by $\rho _{grav}$. The gross mass (energy) $\rho $ will be reduced by self-binding effects to the net mass (energy) density $\rho _m$, 
\beq \rho_m = \rho - \rho _{grav} \; .\eeq
Of course the total energy density $\rho _{tot} = \rho _m +\rho _{grav} $ remains always the same, 
\beq \rho _{tot} = \rho \; , \eeq 
and is only differently distributed between net mass energy density $\rho _m$ and gravitational energy density $\rho _{grav}$.

On   a sphere $S^3$ (later a spherical space) of radius 
\[  R = R_0 f \]
with scaling factor $f$ we consider a mass element $m$. The differential (gross) mass $dm'$ on an infinitesimal strip of width $da$ around the set of all points making  central angle $\alpha \leq \pi $ with respect to $m$ (a 2-sphere) is given by
\[ d m'= \rho V_{S^2}(\alpha )\, da =  \rho V_{S^2}(\alpha )R \, d\alpha   \; , \]
where $ V_{S^2}(\alpha ) = 4 \pi (R \sin \alpha )^2$ is the 2-volume (area) of the 
2-sphere corresponding to $\alpha $. 
\begin{figure}[h!t]
\center{\includegraphics*[scale=0.7]{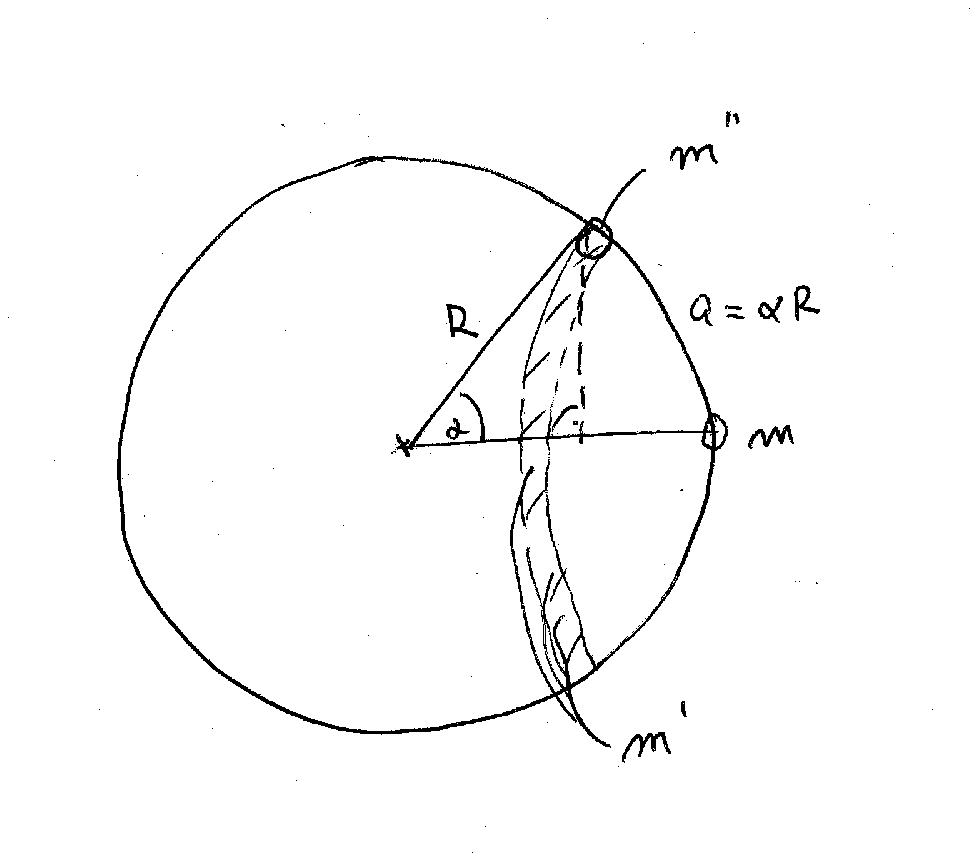}}
\caption{\small  3-sphere with mass elements  $m$, $d m'$ and $dm''$} 
\end{figure}
We assume gravitational forces acting along the geodesic connections between each (well localized) mass element $d m''$ of $dm'$ with $m$ according to Newton's law. The actions along the two complementary geodesic arcs $a=R\alpha $ and $a'=R(2\pi -\alpha )$, connecting $d m''$ and $m$,  lead to inversely oriented forces on $m$. They are  to be subtracted, if we consider both arcs:\footnote{Of course, we might also  consider only the contribution of the main (shortest) arc. Then the coefficients $\beta $ derived below become slightly larger.}
\beq p(dm'') =  G_N (\frac{dm'' m}{a^2} - \frac{dm'' m}{a'^2})  \eeq

During an expansion of the spherical space to $R' = R+dR$ the work to be done is comparable to that  of an expanding lattice. The distance of $m$ and $dm'$ increases by $da = \alpha \, dR $. The contributions  of forces from elements $dm''$   to the work done on $m$ add up to
\[ p(dm') =  4\pi G_N\rho  m R (\sin \alpha)^2 (\frac{1}{\alpha ^2}- \frac{1}{(2\pi -\alpha )^2}) \; da  \; . \]
In  elliptical space $\mathcal{E}^3$ antipodal points of $S^3$ are to be identified. Then the range of $\alpha $ is restricted to $0\leq \alpha \leq\frac{\pi }{2}$ and the complementary arc becomes $(\pi -\alpha )$. Accordingly the last term has to be changed to $\frac{1}{(\pi -\alpha )^2}$.
Integrating over $\alpha $ leads to the work done on $m$ in $S^3$
\beq  d E_m =  4\pi G_N\rho m R \, 4\pi \left( \int _0^{\pi } \frac{\pi -\alpha }{\alpha (2\pi -\alpha )^2} sin^2 \alpha  \; d\alpha \right)  dR \eeq 
and in $\mathcal{E}^3$ to:
\beq    d E_m =  4\pi G_N\rho m R \, \pi \left( \int _0^{\frac{\pi }{2} } \frac{(\pi -2 \alpha )}{\alpha (\pi -\alpha )^2} sin^2 \alpha  \; d\alpha  \right)  dR \eeq
The total mass $M = \rho V_{S^3}(R)$,  with  volume of the 3-sphere 
\[  V_{S^3}(R) = 2 \pi^2 R^3 \; ,\]
remains constant. Thus for $S^3$
\[  dE_m = 8 G_N M m \left( \int _0^{\pi } \frac{\pi -\alpha }{\alpha (2\pi -\alpha )^2} sin^2 \alpha  \; d\alpha \right)\; R^{-2} dR \; , \]
and similarly for elliptical space.
Integration over  the whole   expansion $R\rightarrow \infty$ gives the contribution $E_m$ of $m$ to the self-binding energy  in a 3-sphere of radius $R_0$
\[ E_m =  8 G_N M \frac{m}{R_0}  \int _0^{\pi } \frac{\pi -\alpha }{\alpha (2\pi -\alpha )^2} sin^2 \alpha  \; d\alpha \; .\]

We can now add up over all mass elements $m$ to the total mass  $M$; but then 
 the energy  for each  mass element is counted twice (distance gaining work for  $dm''$ and $m$ is considered from both sides). Cancelling this double count gives the self-binding energy of the mass $M$ homogeneously distributed in an $S^3$ of radius $R$:
\[ E = \frac{1}{2} \,  8 \, G_N  M \frac{ \rho V_{S^3}}{R} \int _0^{\pi } \frac{(\pi -\alpha )}{\alpha (2\pi -\alpha )^2} sin^2 \alpha  \; d\alpha \]

We thus arrive at the gravitational self-binding energy density in the  Newtonian Ansatz for a round $S^3$ 
\beqarr \rho _{grav} &=& \frac{E}{V_{S^3}} = 4G_N \rho ^2 \frac{V_{S^3}}{R}   \int _0^{\pi }  \ldots  \; d\alpha \;  \nonumber \\
 &=& 8 \pi^2 G_N \rho^2 R^2     \int _0^{\pi } \frac{\pi -\alpha }{\alpha (2\pi -\alpha )^2} sin^2 \alpha  \; d\alpha   \eeqarr
For  elliptical space it is 
\beq \rho _{grav} = 2 \pi^2 G_N \rho ^2 R^2  \int _0^{\frac{\pi }{2} } \frac{(\pi -2 \alpha )}{\alpha (\pi -\alpha )^2} sin^2 \alpha  \; d\alpha  \; .  \eeq 

The general form of the gravitational self energy density is 
\beq  \label{self energy density general}  \rho _{grav} = \beta \, G_N \rho ^2 R^2   \eeq
with a  dimensionless coefficient $\beta >0$ which is characteristic for the specific model.

 For studies of dynamical behaviour of cosmological models we can consider {\em the  form   (\ref{self energy density general}) as a  slightly generalized Newtonian approximation of gravitational self-binding energy.} Taking  the correctness of  physical dimensions of formula (\ref{self energy density general}) into account it seems reasonable to express possible modifications of Newton's dynamics in this context by  variations of  $\beta $, the characteristic factor.\footnote{$\beta $ may be affected already by small modifications of the Newton law at scales well beyond the supercluster level. }

For the round sphere we have found
\beq \label{spherical beta}  \beta_{S} =  8 \pi^2 \int _0^{\pi } \frac{(\pi -\alpha )}{\alpha (2\pi -\alpha )^2} sin^2 \alpha  \; d\alpha \;  \approx 6.86 \; ,\eeq
while for the elliptical case (the round projective 3-space)
\beq   \label{elliptical beta}    \beta_{\mathcal{E}} =  2 \pi^2 \int _0^{\frac{\pi }{2} } \frac{(\pi -2 \alpha )}{\alpha (\pi -\alpha )^2} sin^2 \alpha  \; d\alpha   \approx 3.65 \; . \eeq
We shall see in the next section that  specific  values for $\beta$ are decisive for evaluating the qualitative dynamics of the model. 

\section{Consequences for the dynamics of cosmological models}
If we assume, like Fahr e.a., that the vacuum energy $\rho_{vac}$ of cosmology is constituted by gravitational self-binding energy, equ. (\ref{vacuum energy}), we no longer need to hypothesize an agency  like ``dynamical dark energy''  which strongly acts on matter and physical space time,  but is not acted upon. The cosmological constant then turns out as a heuristic device serving  as a formal placeholder which can be used to explore whether it is necessary to extend simple matter models by a vacuum energy term.\footnote{Compare \cite{Earman:Lambda}.}
 If it turns out to be necessary, as recent observational evidence indicates, it requires a more physical explanation. Gravitational self-binding energy may offer a route to the solution of the riddle; at least it seems able to shed new light on it.

Already a rough qualitative consideration shows an interesting fall off behaviour of gravitational vacuum energy in spherical spaces:
\beq \rho_{tot} \sim R^{-3} \; , \quad \quad \rho_{vac} \sim R^{-4}  \eeq
 If one starts close to an equilibrium point characteristic for the Einstein universe,
\beq \label{hyle}  \rho_{vac} = -p = \frac{1}{3} \rho_{tot} \; ,  \quad \quad \rho_{tot } = \rho \; , \eeq
called the {\em hyle condition} for a cosmic fluid in the sequel, a small surplus of mass density will lead to a contraction of the spherical space. Then $\rho_{vac} \sim R^{-4} $ and with it the negative pressure of the vacuum term will increase faster than $ \rho \sim R^{-3}$. This  may, under certain restrictions for the parameters, bring the contraction to a halt and  revert it. Similarly, but conversely, for a fall of mass density below the hyle point the initial expansion may come to a halt,  because the negative pressure falls faster than the total energy density. So we have good reasons to expect, under certain parameter restrictions,  a Lyapunov stable oscillating beviour of spherical space models with gravitational self-binding energy about the Einstein universe (the hyle condition). 

To investigate the case more closely, we have to look at the reduced Raychaudhury equation for the scaling function $f$ of a Robertson Walker solution for the Einstein equation 
\beq \label{Raychaudhury} \frac{f''}{f} = - \frac{4\pi G_N}{3}  (\rho + 3p) \; , \eeq
cf. \cite[A44]{Ellis:83years}, \cite[346]{ONeill}. With (\ref{self energy density general}) we get a nonlinear ordinary differential equation of the form \footnote{In \cite[equ. (5)]{Fischer:1993} this equation has been derived by assuming without further ado that the negative pressure of the Einstein universe derives from gravitational self-energy,  and the qualitiative behaviour has been correctly sketched.}
\beq  \label{Raychaudhury simplified}   f'' = - a_1 f ^{-2}+ a_2 f^{-3} \; . \eeq
The hyle condition ($\rho +3p=0$) may be  normalized to  
\[  f(0)=1, \quad a_1 = a_2 = 1 \; . \]
Obviously the initial condition $f'(0)=0$  leads to a static solution $f\equiv 1$, corresponding to the Einstein universe.

For  initial conditions with small $f'(0)$, an oscillating solution is obtained; for larger initial expansion, the   solution expands monotonically  (figure 2).
\begin{figure}[h!t]
\center{\includegraphics*[scale=0.6]{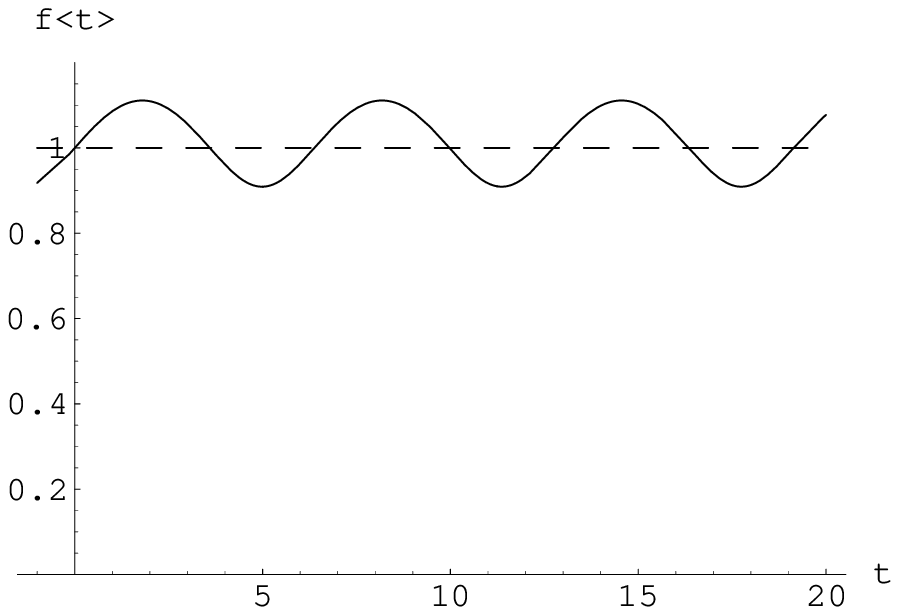}  \quad \includegraphics*[scale=0.6]{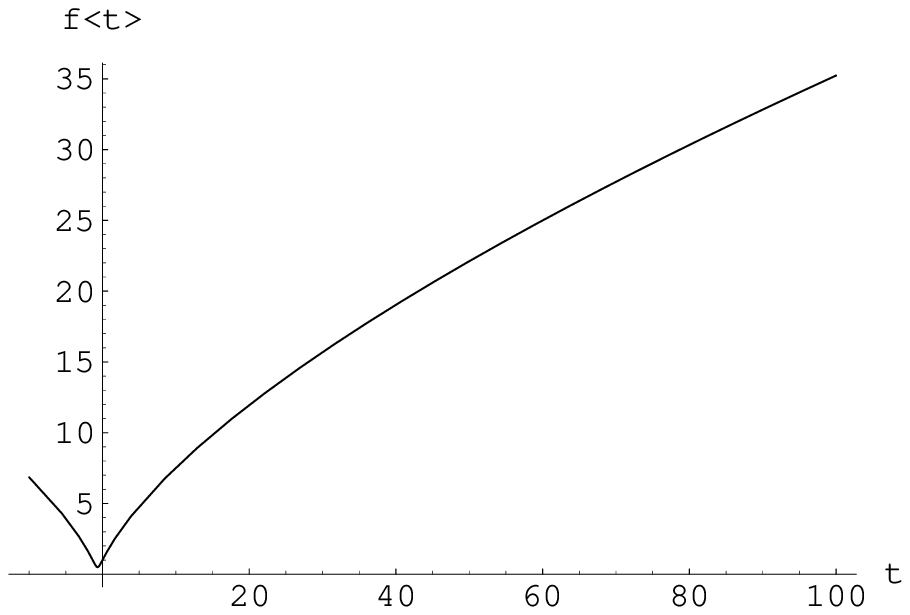}}
\caption{\small Solutions of $ f'' = - a_1 f ^{-2}+ a_2 f^{-3} $ for $a_1= a_2=1$; initial conditions top: $f(0)= 1$, $f'(0)=0$ (dashed), $f'(0)=0.1$ (undashed); bottom:  $f'(0)=1$ } 
\end{figure}

For different coefficients, e.g. $a_2>a_1$,  the oscillatory solutions loose their 
up-down symmetry but may still be periodic (fig. 3). 
\begin{figure}[h!t]
\center{\includegraphics*[scale=0.55]{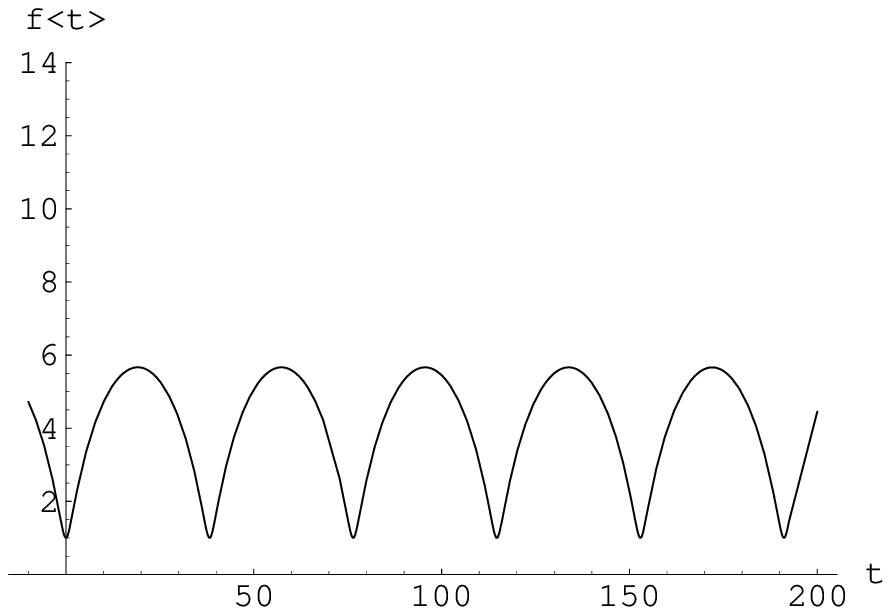} \quad \includegraphics*[scale=0.55]{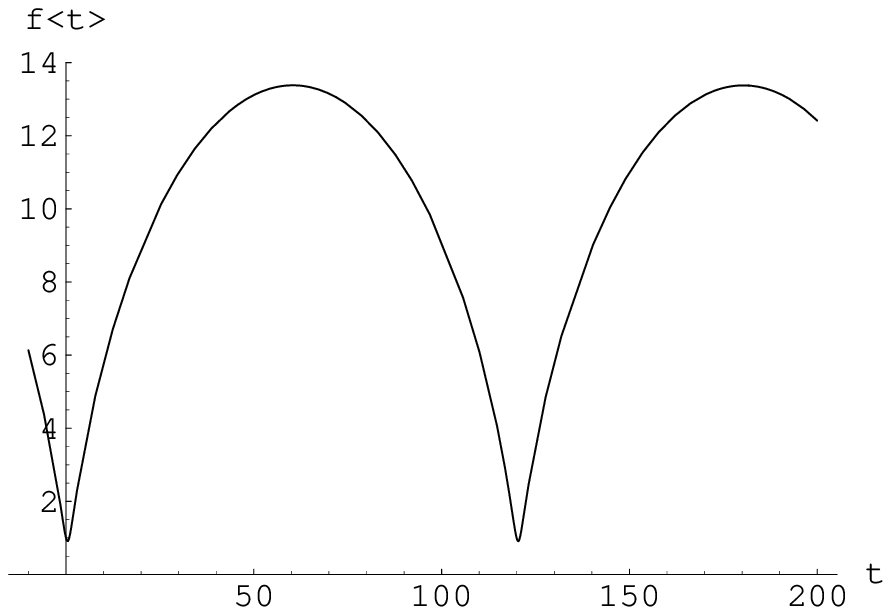} %\quad \includegraphics*[scale=0.5]{abb5}
}
\caption{\small $ f'' = - a_1 f ^{-2}+ a_2 f^{-3} $ for $a_1=1, a_2=1.7$, $f(0)= 1$,  $f'(0)=0$ (left) and $f(0)= 1$, $f'(0)=-0.4$ (right)} 
\end{figure}
The inversion point  for the initial conditions $f(0)= 1, f'(0)=-0.4$  (fig. 3 left) is no cusp, but  a differentiable turning point from contraction to expansion (fig. 4).
\begin{figure}[h!t]
\center{\quad \includegraphics*[scale=0.5]{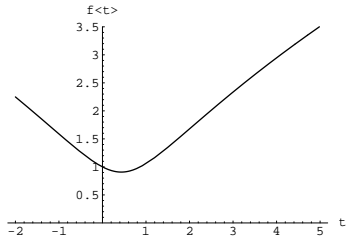}
}
\caption{\small A closer look at the turning point for $ f'' = - a_1 f ^{-2}+ a_2 f^{-3} $ for $a_1=1, a_2=1.7$, $f(0)= 1$, $f'(0)=-0.4$} 
\end{figure}

For increasing ratio $\frac{a_2}{a_1}$ the periods of the oscillations increase. Above a certain bound $\frac{a_2}{a_1} \geq \alpha $, the expansive power of the vacuum pressure prevails; the solution expands monotonically.
Numerical investigations indicate  a bound  $\alpha \approx 1.95 $ for $f'(0)=0$ and  lower  for $f'(0) \neq 0$. Numerical explorations thus indicate a regime of Lyapunov stability about the static solution.

{\em  In this sense, the Einstein universe is theoretically vindicated from a dynamical point of view}. Eddington's famous charge of instability holds for the cosmic constant Ansatz of vacuum energy, but does not hold for the gravitational self energy approach.  Here the Einstein universe reappears as  the static neutral mode of a Lyapunov stable regime of equation (\ref{Raychaudhury simplified}). It remains to be seen, whether this observation may be of  empirical value.  

\section{Application to Einstein-Weyl models}
Contrary to a widely shared opinion, cosmological redshift need not necessarily be the result of a ``true'' space expansion. We cannot exclude  that it may be due to a vacuum loss of photon energy or a higher order gravitational effect. Mathematically the two  physical hypotheses, space expansion and photon energy loss by other reasons,   are interchangeable by using integrable Weyl geometry. The (Weylian)  metric is given, in  {\em gauge}, by a pair $(g, \varphi)$ consisting  of a Lorentz metric $g=(g_{ij})$ and a real differential 1-form $\varphi = \sum \varphi_i dx^i$ (in short $\varphi= (\varphi_i)$). In our context, using  a well known form of the 
Robertson-Walker metric,  they acquire the form
\beq   g: ds^2 = - dt^2 + f(t)^2 \left(\frac{dr^2}{1-kr^2} + r^2 d \Theta ^2 + r^2 \sin^2 \Theta \, d\phi ^2 \right), \;  \varphi = H(t) dt  \eeq 
($k = 0$ or $ \pm 1$). Gauge changes are obtained by rescaling the Riemannian component $g$ of the gauge metric, $\tilde{g} = \Omega^2 (t) \, g,$ and  concomitant {\em gauge transformations} of the differential form $\tilde{\varphi} = \varphi - d \log \Omega $. 
In the framework of Weyl geometry, the  warp (scale) function of Robertson-Walker cosmologies may be ``gauged away''; then the redshift characteristic of the classical warp function is expressed by the Weylian length (scale) connection $\varphi$ only. If cosmological models  in such a gauge (called {\em Hubble gauge} in \cite{Scholz:ModelBuilding}) turn out dynamically and empirically superior to the standard approach, we have to take this as a strong indicator {\em against } the expanding space hypothesis for the Hubble effect (cosmological redshift) and {\em in favour} of the vacuum or field theoretic one.

A class of particularly simple models arises from regauging classical 
Ro\-bert\-son-Walker models with a linear warp function $f(t)= H t$ with  constant $H\geq 0$.\footnote{It may be not by  chance that Fahr e.a.'s {\em economical universe} condition  leads exactly to this type with a linear expansion function. These authors continue to work in the classical (i.e., semi-Riemannian) Robertson-Walker framework.}  A distinguished gauge (Hubble gauge = Weyl gauge) leads to
\beq   g: ds^2 = - dt^2 +  R^2 \left( \frac{dr^2}{1- k r^2} + r^2 d \Theta ^2 + r^2 \sin^2 \Theta \, d\phi ^2 \right) , \quad \varphi = H dt  \; .\eeq 
It indicates a spatial geometry of constant curvature
$  \kappa = k R^{-2} $ and a time independent redshift characteristic with Hubble constant $H$. 

These cosmologies have been termed {\em Weyl universes}, and for $k>0$ {\em 
Einstein-Weyl universes}. Two Weyl universes are isomorphic (in the sense of Weyl geometry) if their parameters (metrical modules) 
\beq \label{zeta} \zeta :=H^{-2} \kappa = H^{-2} R^{-2} \eeq
 coincide. Energy densities are
\[   \rho  = \rho _m + \rho _{vac} \; , \quad \quad  \rho = \Omega  \rho _{crit}\, ,  \;\; \rho_m = \Omega_m  \rho _{crit}\, ,\quad \mbox{etc. } \;\; \]
 with critical density
\beq  \label{rhocrit} \rho _{crit}  =  \frac{3 H^2}{8 \pi G_N }\, . \eeq 
They (have to) satisfy the hyle condition (\ref{hyle}), which means $\Omega _{vac}= \frac{\Omega }{3}$ and $\Omega _m = \frac{2}{3} \Omega $. Moreover, in this model class the metrical parameter is determined by the total energy density:
\beq \zeta = \Omega - 1  \eeq  

The class shows surprisingly good  empirical properties. Supernovae data stand in excellent agreement with positively curved Weyl universes; recent data indicate\footnote{The fit has been improved with respect to \cite{Scholz:ModelBuilding} by  the more recent data of  \cite{Riess_ea:2007}. With a mean square error $\sigma \approx 0.21 $ for the deviation of the model magnitudes from  the data points, the fit of the Einstein-Weyl model is now slightly better than that  of the standard model with $\sigma \approx 0.27 $. The mean standard data error for magnitudes is $\sigma _{dat} \approx 0.24$. }
\[ \zeta = 2.6 \pm 0.4 \, . \]
This fit hints to much higher values for mean mass energy density and for vacuum energy density, $\Omega _m \approx 2.3$ and $\Omega _{vac} \approx 1.2 $, than  accepted at the moment by the majority of cosmologists. Recent weighing of nearby galaxy groups by Ramallah e.a.  indicates, however, the existence of  much higher mass densities in galaxy groups than presently expected, with the consequence that $\Omega _m $ might go up to $\approx 3$ \cite[10]{Lieu:2007}. The final word on mass densities in the universe seems not yet to be spoken. There is no reason to discard the hypothesis of Einstein-Weyl models in cosmology on the basis of presently preferred values for mass densities. 

Vacuum energy density, on the other hand,  behaves much more trustworthy in the new framework.  The standard model of cosmology displays a surprising and even crazy looking shift between mass and vacuum energy densities  during cosmological ``evolution'', which allows relations $\Omega_{vac}\approx 0.75, \, \Omega _m\approx 0.25$ only for a cosmologically short transitory  period \cite{Carroll:Constant}. 
The Weyl geometric models are not affected by such anomalies. Here 
 $\Omega _m$ and $\Omega _{vac}$ remain in a narrow interval corridor. They remain basically constant, with small fluctuations estimated below. It  therefore seems highly interesting  to check, whether the hypothesis of a gravitational
 self-binding energy as origin of vacuum energy is consistent with observational data. 

The generalized Newtonian approximation for self-binding energy in spherical spaces (\ref{self energy density general}) is consistent with the  hyle condition if and only if
\[ \rho _{vac}= \beta_0 G_N \rho ^2 R^2 = \frac{\rho }{3}    \, .\]
Using
 (\ref{rhocrit}), (\ref{zeta}) leads to
\beq  \beta_0 = \frac{8 \pi \, \zeta }{9(\zeta +1)}  \; .\eeq
For   $\zeta \approx 2.6$, determined by the supernovae data,  we find
\[  \beta _0 \approx 2.02 \; . \]  
In the exact Newton Ansatz for the sphere, (\ref{spherical beta}), the gravitational 
self-energy of the spherical Einstein universe lies more than a factor 3 higher, while the elliptical space (the round projective space) comes down to a factor
\[  \frac{a_2}{a_1} \approx \frac{3.65}{2.02} \approx 1.7  \]
in the parameters of the simplified Raychaudhury equation (\ref{Raychaudhury simplified}). The latter constellation lies in the Lyapunov stable regime,  the former does not (see fig. 3).

We conclude: {\em In the  framework of Newtonian  gravitational self-binding energy as source of cosmic vacuum energy the elliptical  Einstein-Weyl universe is dynamically consistent with the supernovae data. The spherical Einstein-Weyl universe is consistent in this sense only if the characteristic coefficient $\beta $ of a Newtonian approximation in equ. (\ref{self energy density general}) is about 40 \% lower than in the exact Newton Ansatz. Otherwise  it leads to an (unboundedly) expanding solution. }

The  time unit $[t]$ in the calculation of fig. 3 is
\[ [t]  = H_0^{-1} \sqrt{\frac{2}{\pi(\zeta +1)}} \; .  \]
For $\zeta \approx 2.6$ that is $[t]\approx 0.4\, H_0^{-1}$. Typical periodicities of solutions are   at the order of magnitude of 10 Hubble times, and the oscillation factor  at the order of magnitude 10 (fig. 3). 

The model displays a very slow pulsation about the static neutral mode with a moderate amplitude. {\em For any observational purpose  the dynamical model  is very well approximated by the corresponding static 
Einstein-Weyl universe.} The redshift component arising from expansion has,  in principle, to be superimposed to the one due to the Hubble form $\varphi=Hdt$;  but  observationally it is negligible. It is intriguing to see how gravitational self energy, already in its Newtonian approximation, is apt to stabilize the geometry of 
Einstein-Weyl universes.

With $\zeta \approx 2.6 ^{+0.3}_{-0.4}$ the first conjugate point of the Einstein-Weyl universe (both for $S^3$ and the $\mathcal{E}^3$) has redshift close to $z\approx 6^{+1.3}_{-0.7} $. The ``equator'' of the sphere corresponds to a redshift $z\approx 1.65 ^{+0.25}_{-0.14}$. If cosmological geometry of the spacelike fibres is elliptic, we should see objects close to the ``equator'' twice, in opposite directions and with slightly different redshifts. A simple empirical test could rely on quasar observations  close to $z\approx 1.6$, which are   exceptionally bright or share exceptional spectral, radio, or X-ray characteristics. They should appear in characteristic  pairs. An empirical test for symmetry constellations of quasars expected in more complicated  spherical spaces could be designed similarly. This would be a continuation, in a new research context, of the search for indicators of a non-trivial topology of the ``universe'' in the large, which has been attempted  in the standard approach exploiting the latter's peculiar view of the cosmic microwave background \cite{Cornish_ea:Topology,Weeks:Topology}.

\section{Conclusion }
This paper contributes to the studies of  gravitational self-binding energy which seem to   offer  a theoretically fruitful, and perhaps even empirically promising, route towards attacking the cosmic vacuum riddle. At the least they open up new perspectives at a theoretical level. It seems remarkable  that already a Newtonian approximation for gravitational binding energy leads to unexpected  dynamical consequences for  traditional cosmological models which have been discarded for a long time. In particular the Einstein universe turns out as a neutral, stationary core state of a Lyapunov stable regime of cosmological models with closed spacelike fibres. In this sense, the Einstein universe comes out vindicated against Eddington's charge of 
instability. A similar idea has been  indicated in  \cite{Fischer:1993}.  Eddington's warning applies, of course, to a cosmic constant   ($\Lambda$-) Ansatz for vacuum energy. 

Our investigation shows how the replacement of the formal-heuristic device of a $\Lambda $-term in the Lagrangian of cosmological models by a more physical hypothesis for vacuum energy may be able to  change the overall dynamical behaviour of Robertson-Walker cosmologies drastically. 
The  contributions of mass and gravitational self-energy to the total energy momentum seems to behave, under certain not unrealistic restrictions, like a 
self-stabilizing fluid. This is the justification for the word ``hyle'' condition of the whole stabilized system (\ref{hyle}).\footnote{ {\em hyle} $\sim $  (Greek) original substance. The  original substance of Thales, the oldest Ionian natural philosopher known by name,  was a kind of primordial fluid (``water''). 
In \cite{Scholz:Variationsansatz}  a much more  formal  isentropic fluid Ansatz has been studied. It also led to stabilizing conditions. }
Expectations of this kind  seem to have been around among the first generation of relativists. Tullio Levi-Civita talked about the possibility of ``real fluids'' with negative pressure \cite[359, 429]{Levi-Civita:1926}. This formulation made sense only with reference to cosmological considerations. At the end of the book he discussed the Einstein universe; but there seem to be no further notes on this question.\footnote{I thank R. Tazzioli, coeditor of the Levi-Civita papers, for this information.}

Already this result  underpins the  demand that a {\em physical concept} of the cosmic vacuum (in contrast to a purely formal-heuristic one)  ought to 
 take  actions in both directions  into account, from vacuum on  spacetime and matter, and from matter on vacuum. This should cast doubts on  a ``dynamical dark energy'' as an agency which violates this basic principle. 

So far our conclusions refer to primarily methodological questions. But there is more to resume. Our investigation  has shown that the observational data on supernovae magnitudes are consistent with a  Weyl geometric approach  in a slightly modified Newtonian approximation with characteristic coefficient $\beta $ about $60 \; \%$ of the exact Newton Ansatz, or smaller. An 
   {\em  unmodified} Newton An\-satz  for gravitational self-binding energy is consistent with the empirical data (if and) only if the spatial geometry is of a more complex topology than the sphere.  A closer look at  quasar data may be able to decide  whether typical symmetry constellations of quasars, or other exceptional astronomical objects,  about the `cosmic equator' close to $z\approx 1.65$  do occur  empirically. Most easily testable will be the  simple antipodal symmetry of the elliptic case,   $\mathcal{E}^3$. 
\vspace{20mm}
\small 

%\newpage
\bibliographystyle{apsr}
\bibliography{a_litfile}

\end{document}